\documentclass[preprint,aps]{revtex4}
\usepackage{graphicx}
\usepackage{booktabs}
\usepackage[usenames, dvipsnames]{color}
\usepackage{multirow}
\usepackage{eurosym}
\usepackage{setspace}
\usepackage{amsmath}
\usepackage{mathtools}
\usepackage[export]{adjustbox}
\usepackage{graphicx}
\usepackage{epstopdf}
\usepackage{rotating}
\usepackage{color}
\usepackage[ pdftex, plainpages = false, pdfpagelabels, 
                 pdfpagelayout = useoutlines,
                 bookmarks,
                 bookmarksopen = true,
                 bookmarksnumbered = true,
                 breaklinks = true,
                 linktocpage=all,
                 pagebackref=false,
                 colorlinks = true,
                 linkcolor = BrickRed,
                 urlcolor  = blue,
                 citecolor = BrickRed,
                 anchorcolor = green,
                 hyperindex = true,
                 hyperfigures
                 ]{hyperref}

\begin{document}

\title{\href{http://www.necsi.edu/research/psychosocial/diffreddit/}{How Do People Differ? A Social Media Approach}} 
\date{\today}  
\author{Vincent Wong and \href{http://necsi.edu/faculty/bar-yam.html}{Yaneer Bar-Yam}}
\affiliation{\href{http://www.necsi.edu}{New England Complex Systems Institute} \\ 
210 Broadway, Suite 101, Cambridge, MA 02139, USA}

\begin{abstract}
Research from a variety of fields including psychology and linguistics have found correlations and patterns in personal attributes and behavior, but efforts to understand the broader heterogeneity in human behavior have not yet integrated these approaches and perspectives with a cohesive methodology. Here we extract patterns in behavior and relate those patterns together in a high-dimensional picture. We use dimension reduction to analyze word usage in text data from the online discussion platform Reddit. We find that pronouns can be used to characterize the space of the two most prominent dimensions that capture the greatest differences in word usage, even though pronouns were not included in the determination of those dimensions. These patterns overlap with patterns of topics of discussion to reveal relationships between pronouns and topics that can describe the user population. This analysis corroborates findings from past research that have identified word use differences across populations and synthesizes them relative to one another. We believe this is a step toward understanding how differences between people are related to each other. 
\end{abstract}

\maketitle

\section{Introduction}

Since the advent of social media, abundant public data has been produced, and there has been a concerted research effort to understand the people behind the data. Previous studies have adopted different perspectives from disparate fields, including Natural Language Processing (NLP) and personality psychology. Work in sentiment analysis focuses on accurately extrapolating emotional content from text \cite{PennebakerMehl, YangCui, DoddsHarris, Bertrand, LewenbergBachrach, SchwartzUngar}. Other NLP methods involve algorithms that determine topics of conversation \cite{ZhaoJiang, XiangXu} or demographics from language \cite{PietroVolkova, TausczikPennebaker, VolkovaVanDurme}. On the other hand, approaches from personality psychology incorporate predetermined personality traits to obtain correlations between word usage, demographics such as gender, and personality types \cite{GoldbergBig5, DigmanBig5, GoldbergPheno, NormanTaxonomy, opendataKosinski, CareyBrucks, FarnadiSitarman, SchwartzUngar}. A number of studies have used social media data including Facebook and Twitter \cite{FarnadiSitarman, opendataKosinski, ParkSchwartz, LamiotteKosinski}. Furthermore, research outside of these fields uses social media data to understand the dynamic behaviors of users \cite{MillerTweetStream, BachrachGraepel} or to detect medical conditions in individuals, including mental illnesses \cite{PietroEchstaedt,CoppersmithDredze} and heart disease \cite{EichstaedtSchwartz}. 

A challenge that remains unaddressed is the connection between linguistics and personality beyond specific correlations of predefined variables. As a first step in this direction, we use multidimensional scaling (MDS) to investigate differences in individuals by characterizing the high dimensional space of their word usage. We calculate vectors of word counts (feature vectors) and reduce the dimensionality of the space of these vectors to observe patterns. The most important dimensions best capture the differences between individuals as combinations of word usage choices. We investigate content and characteristics of user posts to gain an understanding of what the reduced dimensions reveal about the behavior of users. This data-driven approach is unsupervised and requires no prior assumptions about the axes of personality traits. We analyze data from Reddit, the content rating and discussion website. 

\begin{figure}[h!]	
\begin{center}							
\begin{minipage}[t]{0.5\textwidth}
\includegraphics[width=1.0\textwidth, valign=t]{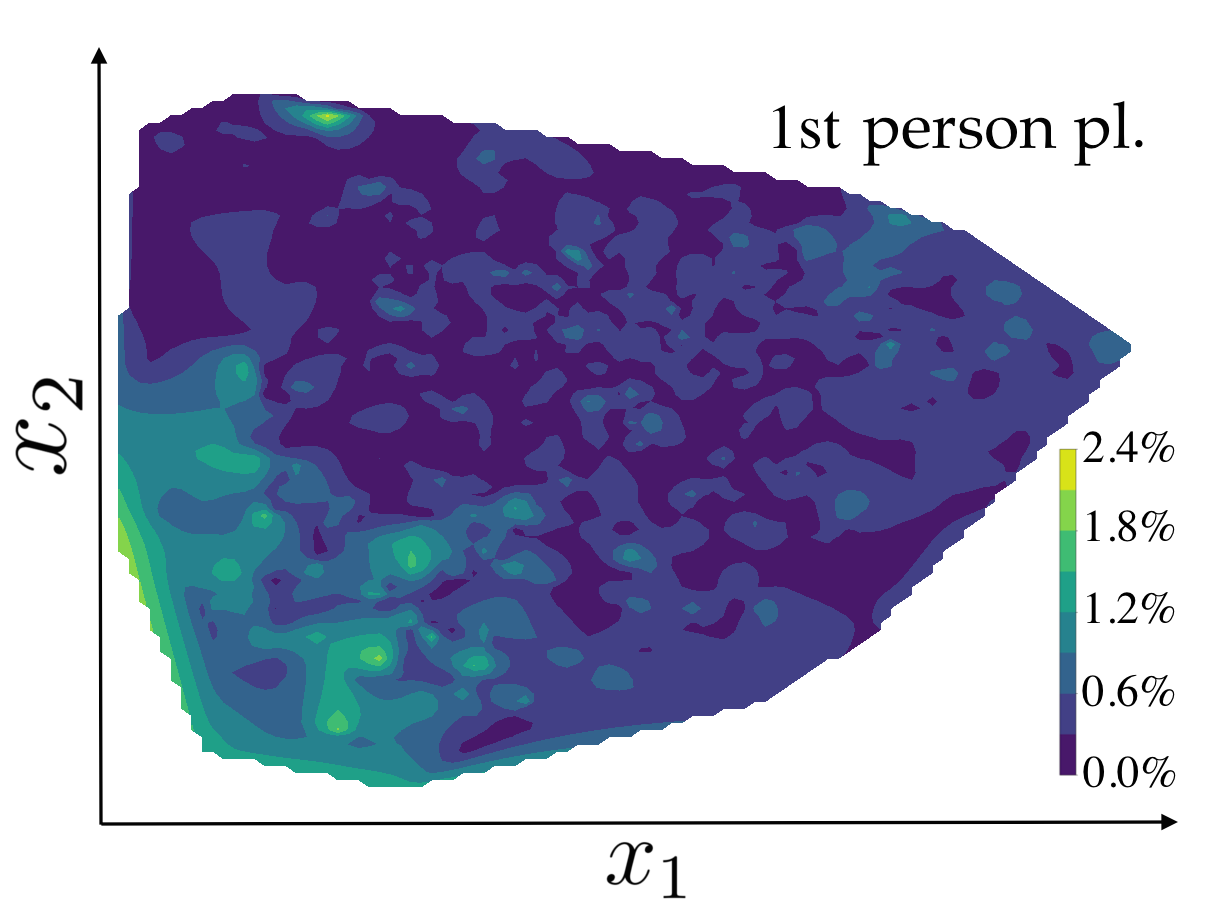}
\end{minipage}

\caption{\label{fig:1}\textbf{Differences in user pronoun use} plotted in the space of the two most prominent dimensions in which individuals differ from each other in word usage ($x_1,x_2$). The region that is colored is the area of the space that is populated by individuals. Color represents the percent of each user's dictionary that is first person plural pronouns (`we',`us',`our',`ours',`ourselves', etc). Areas in the bottom left have a much higher use of these pronouns. Pronouns were not used in the determination of the prominent dimensions of variation so the variation in pronoun use is linked to other word use variations.
}
\end{center}
\end{figure}

Our analysis finds similar correlations as previous research, provides new insights, and relates them to each other. Fig. \ref{fig:1} shows the set of users plotted in the two most important dimensions. Colors show 1st person pronoun use as an example of patterns of differences among them. By looking at multiple ways individuals differ in word usage across these two dimensions, we can identify the codependencies among different behavioral traits, including those that may be associated with personality \cite{NormanTaxonomy, opendataKosinski, CareyBrucks, FarnadiSitarman, SchwartzUngar}. This is a preliminary study in using this type of analysis on social media data. 

\section{Methods}

From a dataset of Reddit posts, we obtained a sample of 1500 users who had made posts or comments at least 50 times in the month of January 2015 \cite{Reddit}. For each user, we created a user dictionary consisting of the counts of each word they used across all of their posts in that month. We determined words by space delineation, and we removed all punctuation as well as Reddit-specific text formatting (e.g. the use of asterisks to make text bold). Common words which primarily serve grammatical function and do not contribute to topical content---known as stop words in computational linguistics---were removed from the dataset (see Appendix A). We construct ``feature vectors" for each user where each dimension corresponds to a word in the total dictionary of words used by all users, and nonzero entries are counts of words from each user dictionary. 

We apply Multidimensional Scaling (MDS) to the set of feature vectors using Torgerson-Gower scaling. We generate a matrix of Euclidean distances between user vectors pairwise for all users, resulting in an $N\times N$ matrix where N is the number of users and the diagonal of the matrix is zero. We plot the users in the space of the eigenvectors with the largest eigenvalues. These composite dimensions can be considered to best characterize the users across the space of word usage.

When plotting users this way, bots were revealed because of their strong differentiation from human users. We removed bots from manually selected regions that show extreme separation from other users. We confirmed that they were bots by inspection. We used the resulting cropped list of 1462 users for our analysis. 

This method is similar to the factor analysis used in identifying the Five Factor model in personality psychology \cite{McCraeJohn, DigmanBig5, GoldbergBig5, GoldbergPheno}. Unlike work done in the framework of that model, the data used here does not restrict itself to a predetermined set of adjectives. It differs also in the use of social media data rather than survey data. 

To investigate the space of users in the reduced dimensions, we compiled statistical measures on each user's dictionaries. We use the following measures: dictionary entropy, unique word count, and percent contribution of pronouns by their person, gender, and number (first-, second-, and third-person singular and first- and third-person plural). 

We calculated the dictionary entropy as 

\begin{equation}
S_u = -\sum_w{\frac{n_w}{\sum_w{n_w}}\log(\frac{n_w}{\sum_w{n_w}})} 
\end{equation}

\noindent where $w$ is an index labeling each word and $n_w$ is the count of that word in the user dictionary. The dictionary entropy can be interpreted as the word diversity. An investigation of raw word counts yielded similar results. 

To determine the topics discussed in the data, we use Latent Dirichlet Allocation (LDA) because it maps directly to the bag-of-words approach used for feature vectors. LDA assumes that each user feature vector (also called a ``document") is generated probabilistically from a distribution of topics, and attempts to derive these topics. The same word can appear in multiple topics. We used collapsed Gibbs sampling to redistribute words to topics given their presence in documents. We generated 100 topics of 40 words each using user feature vectors as documents over 1000 iterations of LDA, with stop words removed. We calculate the percent contribution of a topic to each user's dictionary based on what words appear in topics. We labeled topics based on their content for ease of reference where possible. 

\section{Results}

\begin{figure}[ht!]	
\begin{center}	
\begin{minipage}[t]{0.30\textwidth}
A\includegraphics[width=1.0\textwidth, valign=t]{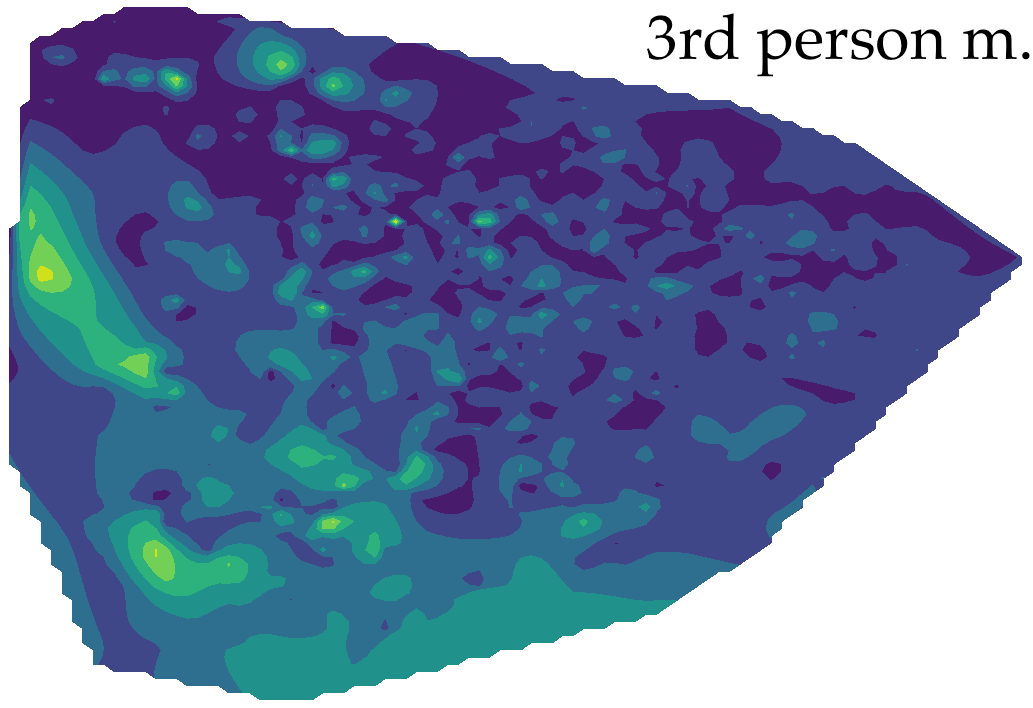}
\end{minipage}\hspace{0.5cm}
\begin{minipage}[t]{0.30\textwidth}
B\includegraphics[width=1.0\textwidth, valign=t]{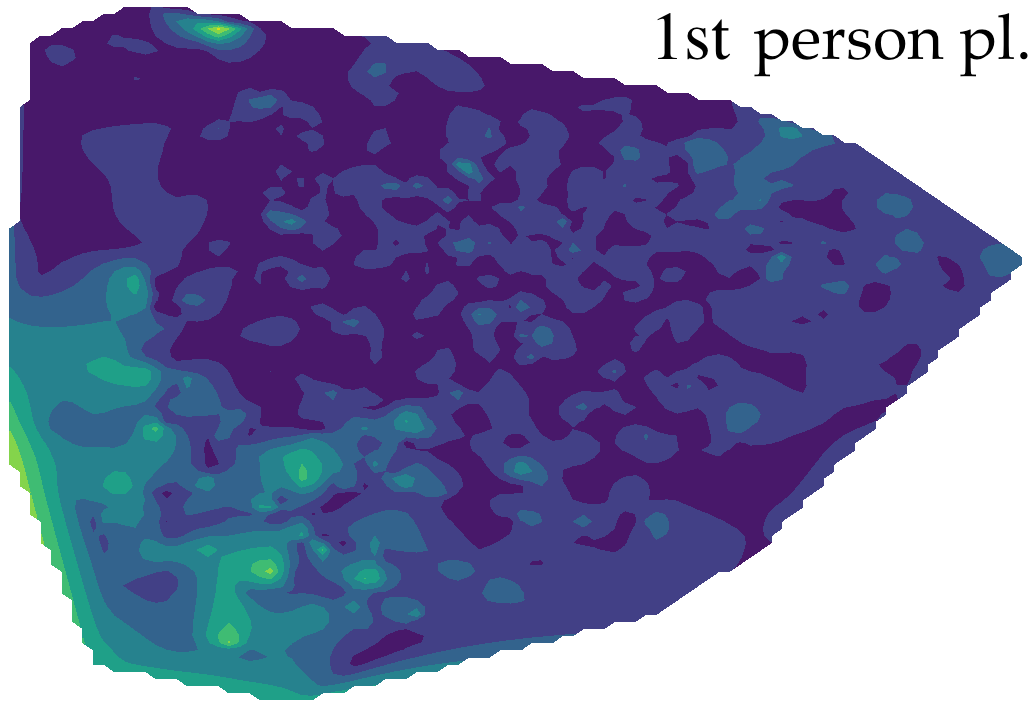}
\end{minipage}\hspace{0.5cm}
\begin{minipage}[t]{0.30\textwidth}
C\includegraphics[width=1.0\textwidth, valign=t]{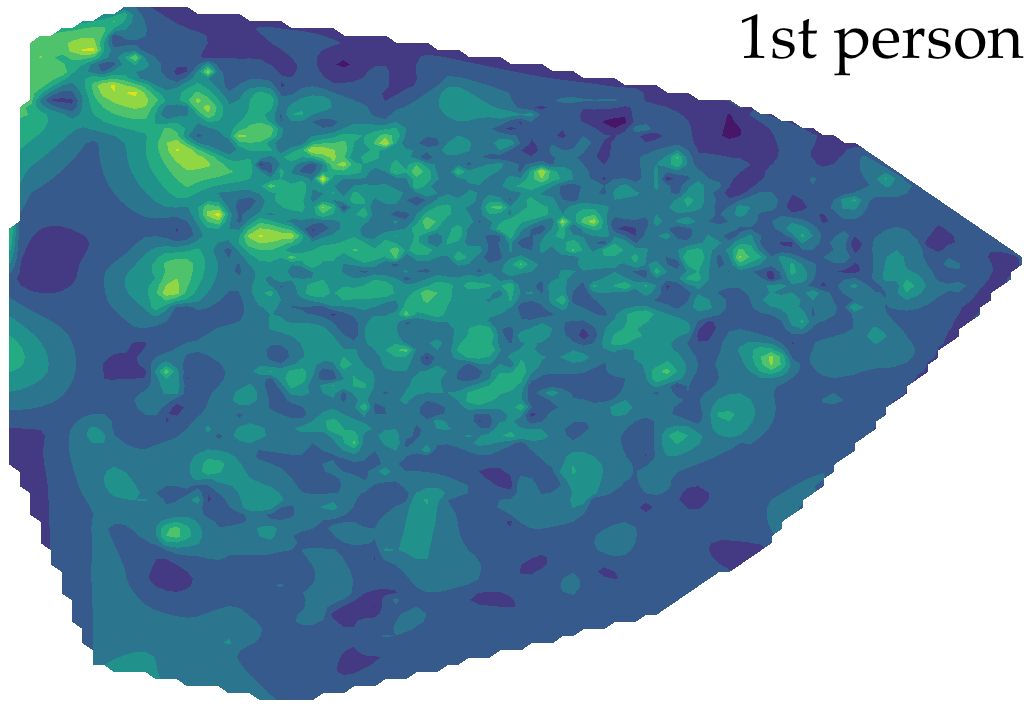}
\end{minipage}
\begin{minipage}[t]{0.30\textwidth}
D\includegraphics[width=1.0\textwidth, valign=t]{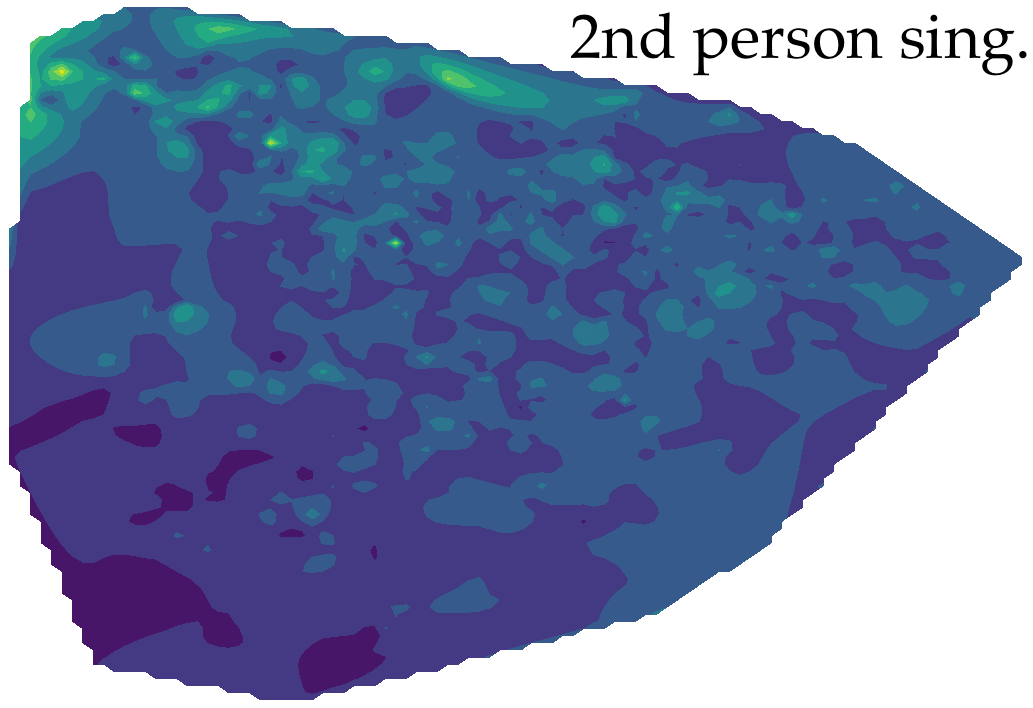}
\end{minipage}\hspace{0.5cm}
\begin{minipage}[t]{0.30\textwidth}
E\includegraphics[width=1.0\textwidth, valign=t]{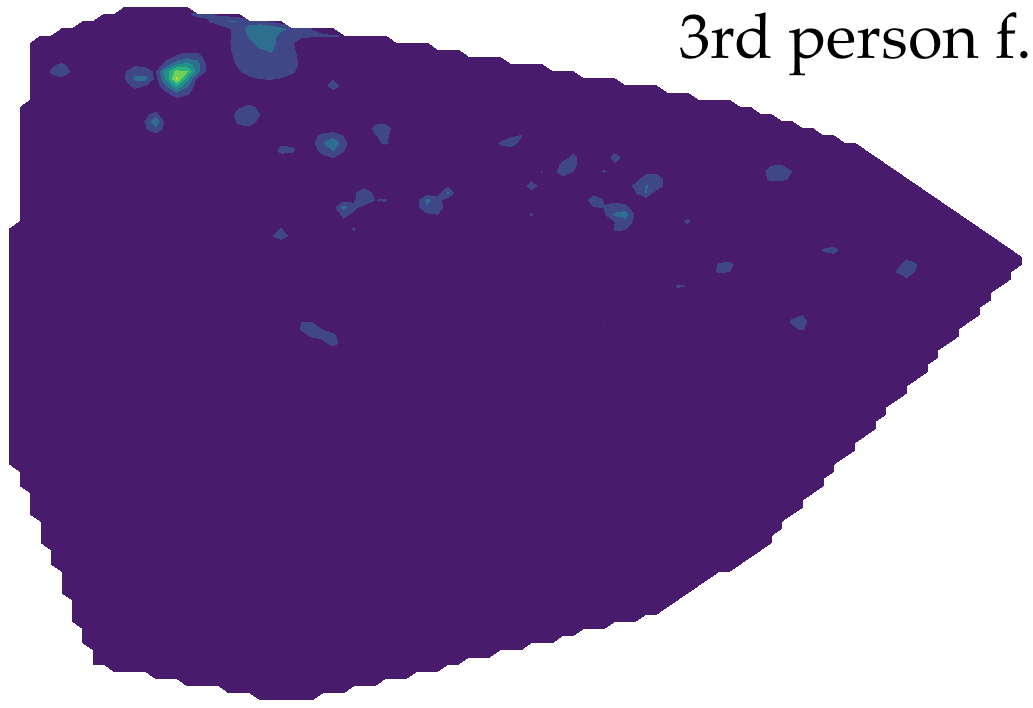}
\end{minipage}\hspace{0.5cm}
\begin{minipage}[t]{0.30\textwidth}
F\includegraphics[width=1.0\textwidth, valign=t]{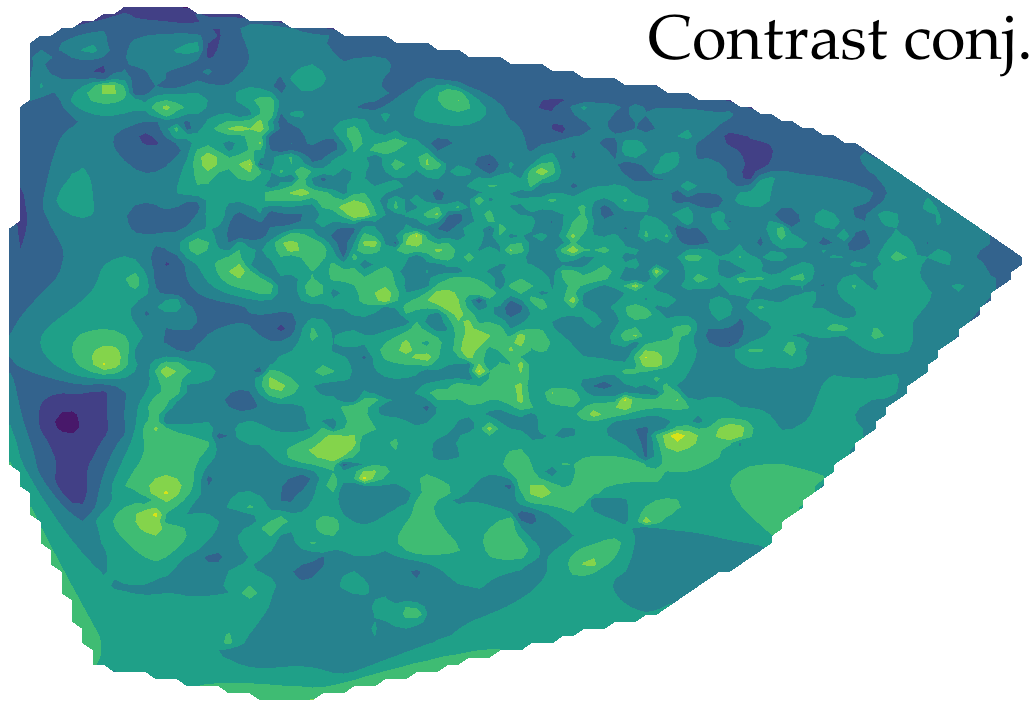}
\end{minipage}
\caption{\label{fig:3}\textbf{Word use metrics plotted in the space of the two most prominent dimensions of variation among individuals.} Each plot shows the same region of space colored by a different metric. Areas of the space with higher levels of usage of pronouns or other words are shown by lighter colors (color bar as in Fig. 1). A. Male third-person pronouns (`he', `his', `him', `himself'). B. First person plural pronouns (`we', `us', `our', `ourselves'). C. First person pronouns (`I',`me',`myself',`my'). D. Second person singular pronouns (`you',`your',`yours',`yourself'). E. Female third-person pronouns (`she',`her',`hers',`herself'). Third person female pronouns are underrepresented in Reddit because the site is male-dominated. F. An example of the analysis of another category of words, here contrasting conjunctions (`but',`though',`however'). 
}
\end{center}
\end{figure}

We constructed a reduced dimensional representation of word usage that captured the primary variations across user dictionaries excluding common words known as ``stop words" (see Methods). The locations of users in the two most prominent dimensions of variation are shown in Fig. \ref{fig:3}. We investigated the individual differences in use of pronoun types which may have implications for variations in personality \cite{NormanTaxonomy, opendataKosinski, CareyBrucks, FarnadiSitarman, SchwartzUngar}. The colors in the first five panels show the percent of types of pronouns in user dictionaries. Since we are constructing our understanding directly from data, even elementary observations are important. We can see that people have significant differences in their use of pronouns. Moreover, individuals in different areas of the reduced space tend to have similar pronoun use. The bottom and left regions have a greater incidence of third-person male pronouns (`he',`his',`himself'). The lower left itself has greater incidence of first-person plural pronouns (`we',`ours',`ourselves'). The middle and upper left areas have a greater use of singular first-person pronouns (`me',`myself',`I'). The top and top left has greater use of second-person pronouns (`you', `yourself'). The few users with high use of third-person female pronouns (`she',`her',`herself') are primarily near the top. 

The observation that pronoun use varies systematically with the dimensions that capture the greatest variance has important implications for our understanding of individual differences. In particular, the primary ways in which individuals differ include significant differences in pronoun use. It is worth noting that pronouns were included among the stop words and were removed from the dimensional reduction process, so the differences in pronoun use were not included in determining these dimensions. The observation of systematic variation implies that the other words people use are linked to pronoun use across the dimensions. 

The spatial structures of the variations reflect not just individual or correlational changes in pronoun use, but also the codependency among these usages. For example, first-person plural pronouns (Fig. \ref{fig:3}B) are high in a subset of areas of third-person male pronouns (Fig. \ref{fig:3}A). Third-person female pronouns (Fig. \ref{fig:3}E) appear to be localized in a subset of areas of high second-person singular pronoun use (Fig. \ref{fig:3}D). Third-person male and first-person plural pronouns (Fig. \ref{fig:3}A and B) occur in areas distinct from second-person singular and third-person female pronouns (Fig. \ref{fig:3}D and E). However, the area of high first person pronoun use (Fig. \ref{fig:3}C) overlaps with both groups but is distinct from them. 

In the last panel of Fig. \ref{fig:3}, we show the use of contrasting conjunctions (`but', `though', `however'). Contrasting conjunctions are widely distributed across two dimensions. However, there is a small patch on the left just below center where usage is substantially lower. Correlations between the two primary dimensions of variation and the measures of contrasting conjunctions would not reveal this behavior. They would fail to capture the spatial variation of the vertical dimension in the middle left of Fig. \ref{fig:3}F because of aggregation across the horizontal dimension as well as the up-down-up variation of the left region. They would also fail to capture the variations in the horizontal dimension due to aggregation in the vertical. The highly nonlinear variations across the space would not be captured in correlations in either dimension. 

\begin{figure}[ht!]	
\begin{center}					
\begin{minipage}[t]{0.49\textwidth}
A \includegraphics[width=1.0\textwidth, valign=t]{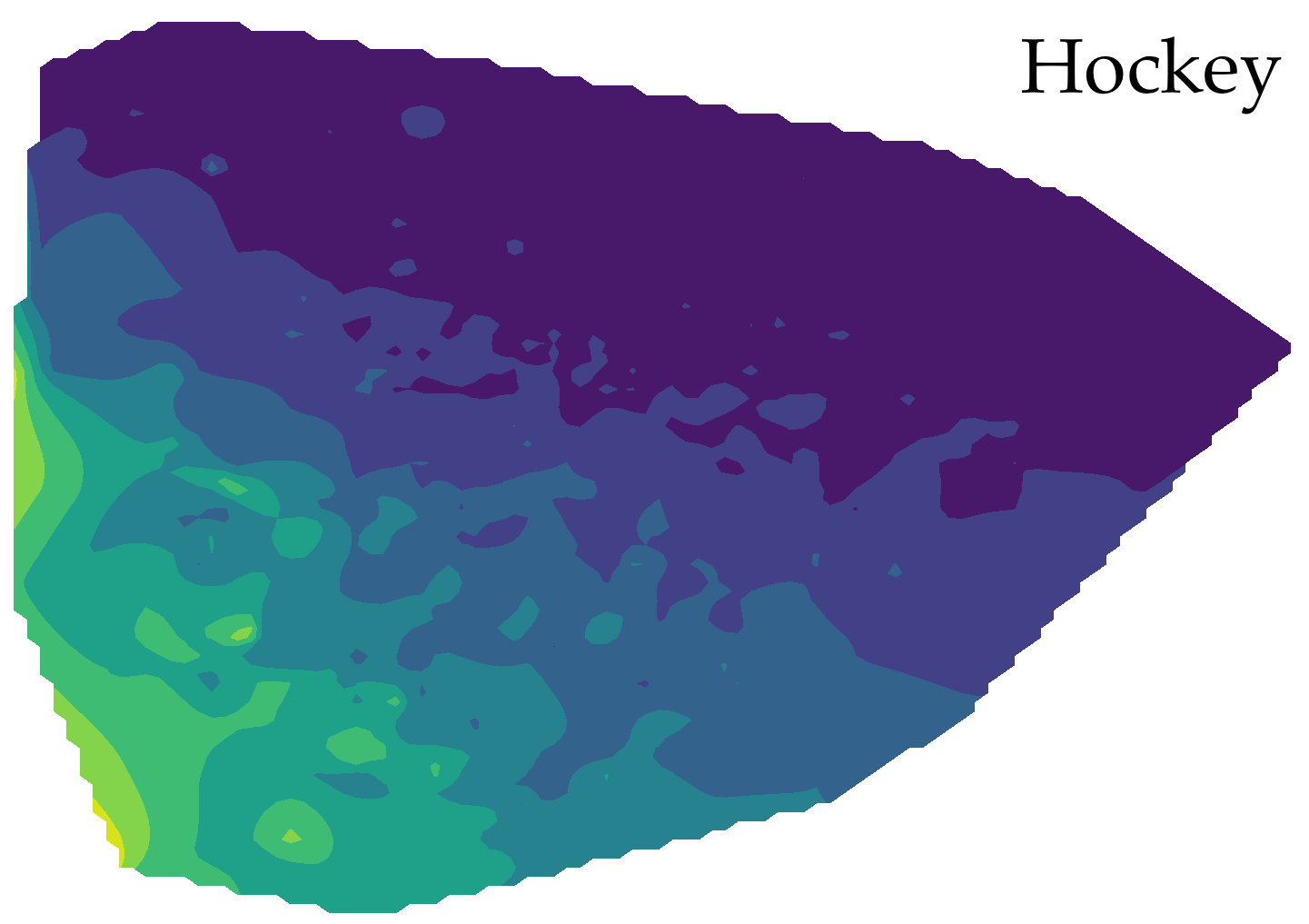}
\end{minipage}
\begin{minipage}[t]{0.49\textwidth}
B \includegraphics[width=1.0\textwidth, valign=t]{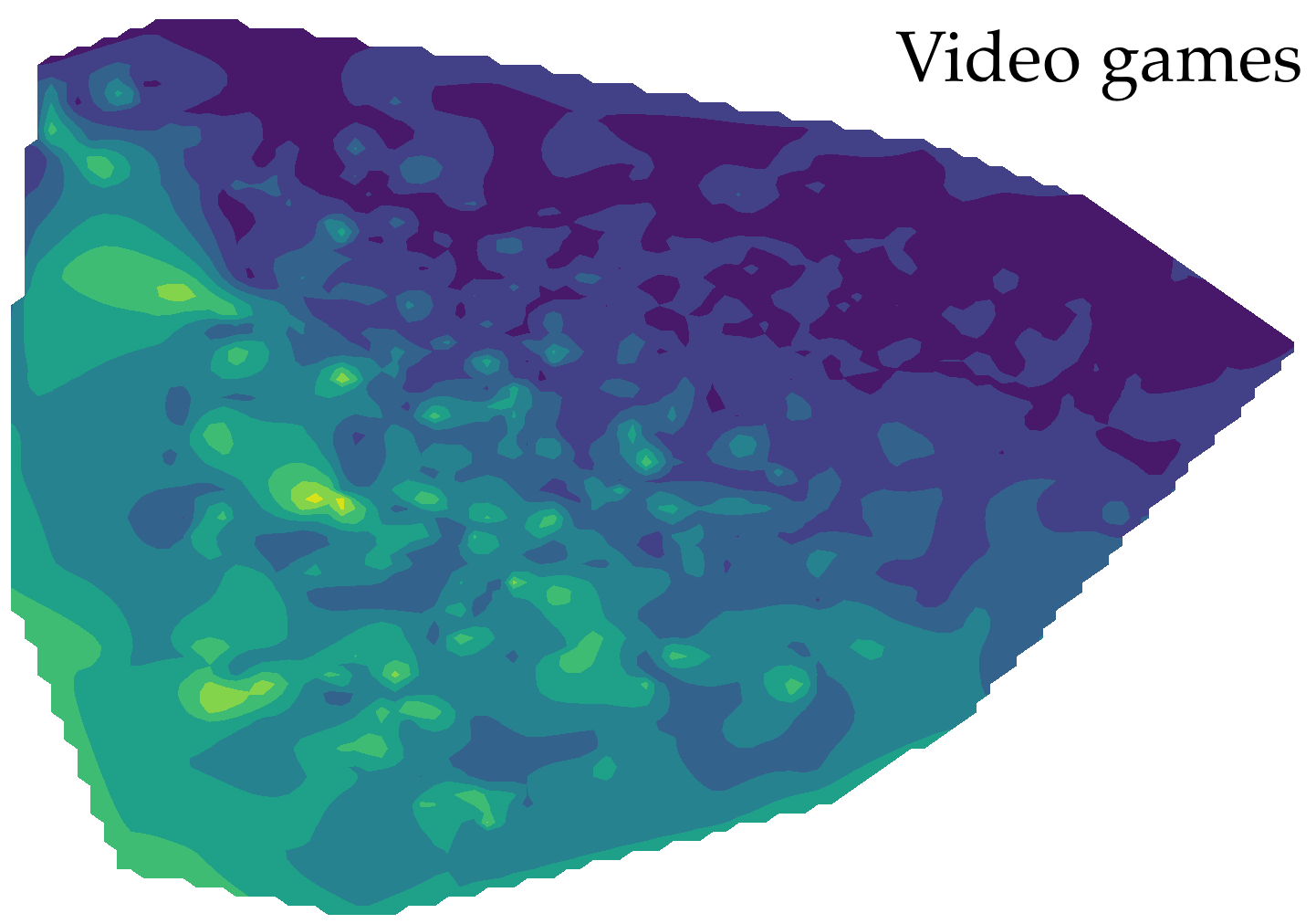}
\end{minipage}
\begin{minipage}[t]{0.49\textwidth}
C  \includegraphics[width=1.0\textwidth, valign=t]{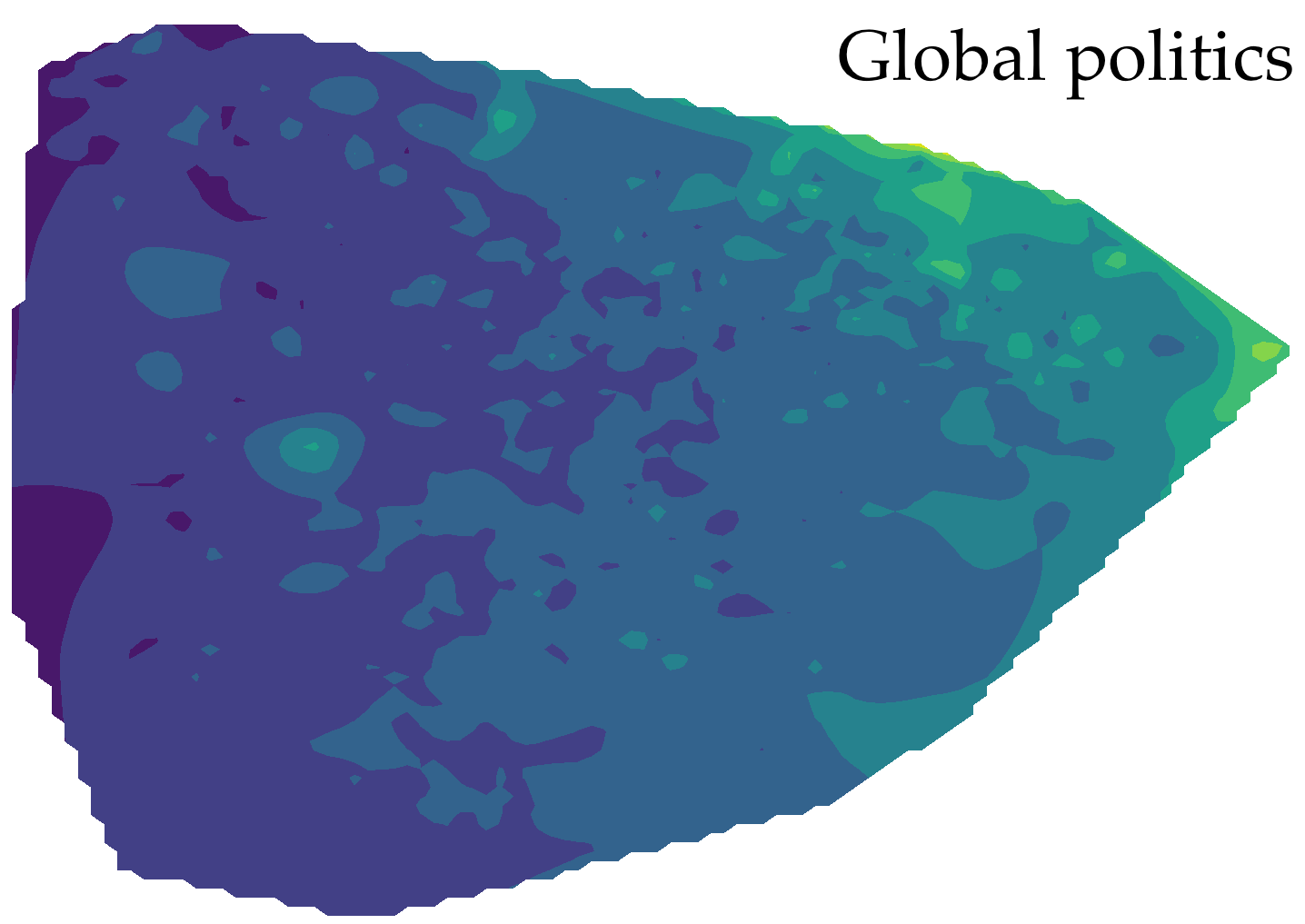}
\end{minipage}
\begin{minipage}[t]{0.49\textwidth}
D \includegraphics[width=1.0\textwidth, valign=t]{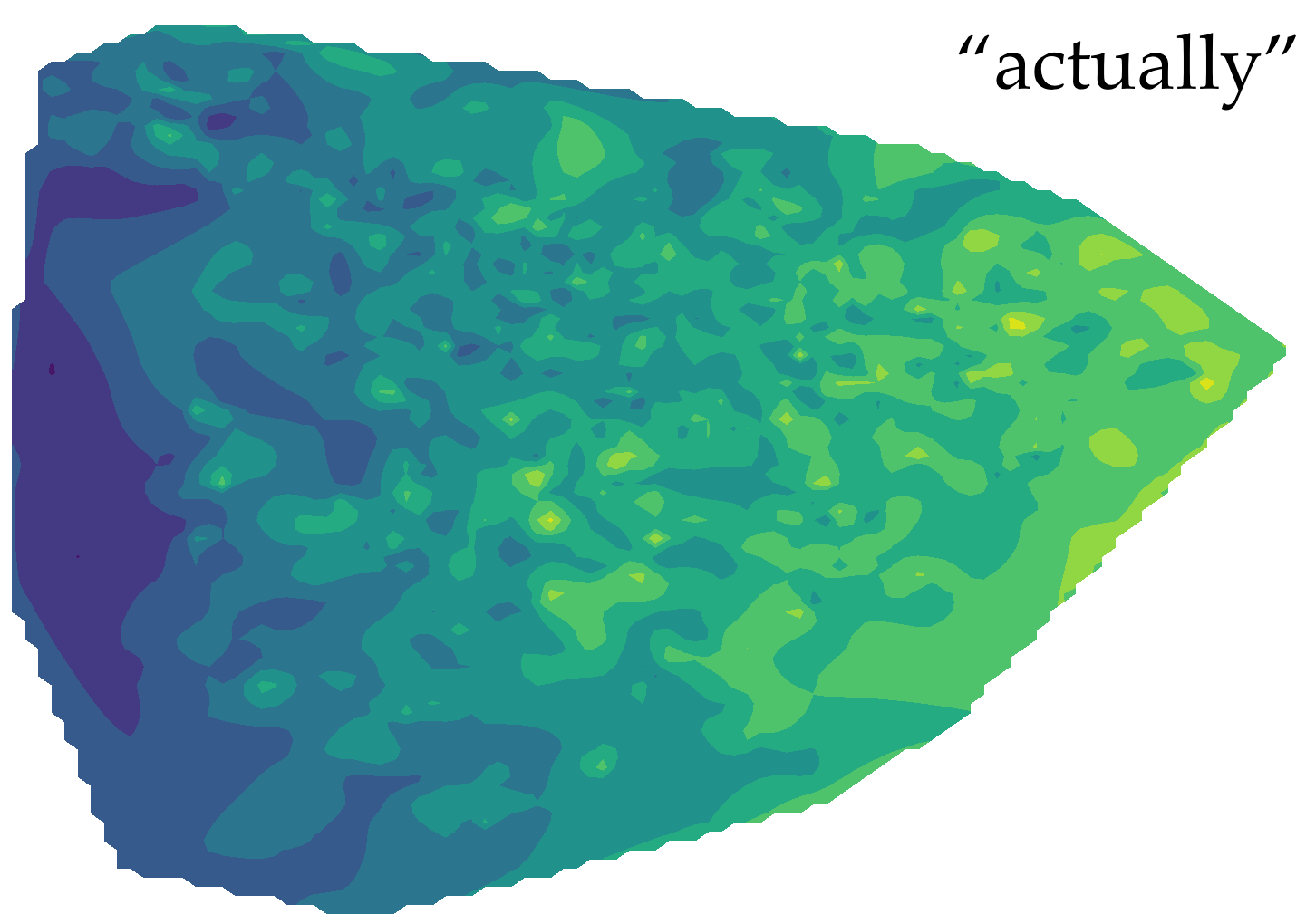}
\end{minipage}
\caption{\label{fig:4}\textbf{Topics of discussion in the space of the two most prominent dimensions of word usage differences} A. Hockey-related words (`NHL', `hockey', ). B. Video game related words (`game',`enemy',`kill',`points'). Both A and B exemplify the way that many of the sport/game related topics extracted with LDA coincide in the population. C. Global politics (`world',`politics',`government',`money') D. An assortment of common words determined by LDA to be a distinct topic labeled by the word ``actually." This analysis shows that this word group is meaningful even if it is non-trivial to determine the concept underpinning it. }
\end{center}
\end{figure}

We investigated the words people use (not including stop words) and identified topics of their conversation and how they vary across users. We used Latent Dirichlet Allocation to extract 100 topics of conversation with 40 words associated with each topic (see Methods). We found the percent contribution of each topic to each user dictionary. The percent contribution of topics across users in the two most prominent dimensions of word usage variation is shown in Fig. \ref{fig:4} for each of four example topics named by inspection of representative words for that topic. The ``hockey'' topic contains words related to the sport of hockey, such as `hockey', `nhl', `puck', `ice', `bruins', and `canucks'. The ``video games'' topic has words such as `damage', `enemy', `items', `attack', and `armor'. The topic ``global politics'' is characterized by words related to interactions between states at the global scale, such as `war', `religion', `people', `money', and `government'. The fourth topic, ``actually'', is labeled by a prominent word from the topic because a clear identifier for the topic was not readily apparent, and it contains the words `actually', `very', `more', `anything', and `lot'. (Full topics in Appendix B)

The topic ``hockey'' occurs more frequently in the bottom left area of the reduced dimensional space (Fig. \ref{fig:4}A). Words related to video games occur in the bottom and left half (Fig. \ref{fig:4}B). Words related to global politics occur in the dictionaries of users in the rightmost area (Fig. \ref{fig:4}C). The topic labeled by the word ``actually'' is predominantly found in the users toward the right region. 

The bottom and left areas that have a greater use of words from sport- and game-related topics also corresponds with a greater use of male third-person pronouns (Fig. \ref{fig:3}A). This may indicate that people who discuss sports and games are also the people who refer to men most often, which may be due to the fact that both sports and video games are male-dominated subjects. In addition, these users have a greater use of first-person plural pronouns (Fig. \ref{fig:3}B), which may be linked to a greater incidence of discussions about teamwork. 

\begin{figure}[ht!]	
\begin{center}							
\begin{minipage}[t]{0.4\textwidth}
\includegraphics[width=1.0\textwidth, valign=t]{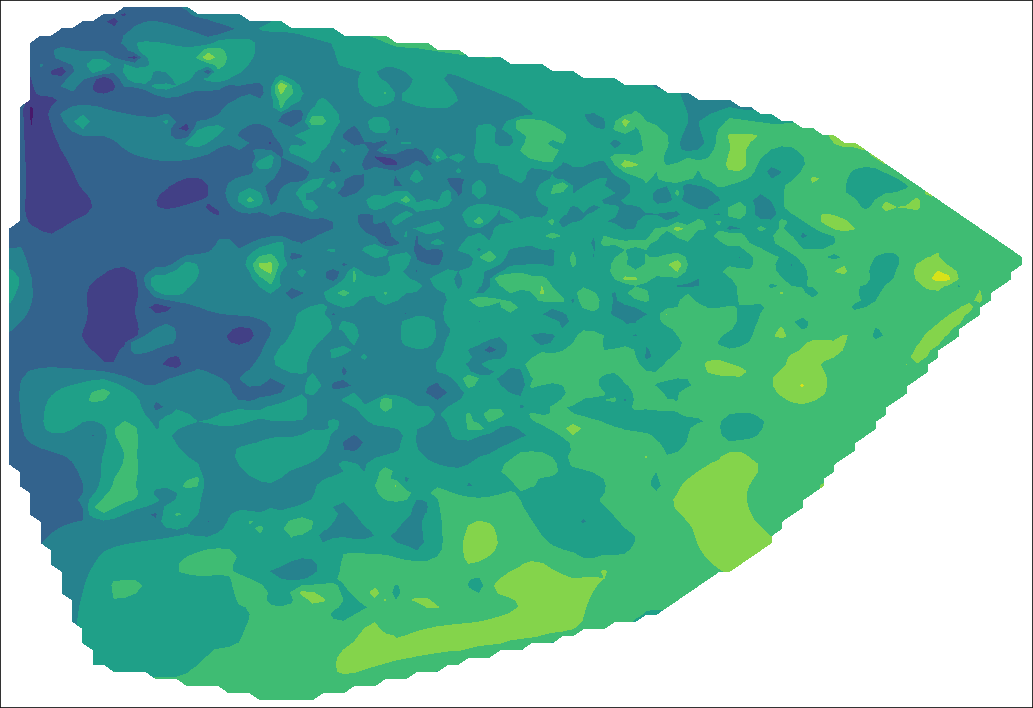}
\end{minipage}
\caption{\label{fig:2}\textbf{Entropy of user dictionaries} plotted as a function of the two most prominent dimensions of individual differences. 
}
\end{center}
\end{figure}

Finally, in Fig. \ref{fig:2} we show the entropy of user dictionaries as a function of the two most prominent dimensions of word usage variation. The entropy is higher in the lower right region of the reduced space, indicating that those users have more varied dictionaries. 

All results shown in Fig. \ref{fig:3} - \ref{fig:2} are statistically significant with the exception of the use of third-person female pronouns in Fig. \ref{fig:3}E. The statistical significance was determined by randomizing the metric values across users and comparing this to the original results with the null hypothesis that the results are unrelated to the position in the reduced space. We calculate the distribution of variances from a sliding window across the interpolated space and applied an independent t-test to compare the variances between the random and original measures. We find a $p < 0.001$ for all measures investigated, except third-person female pronouns which failed to reject the null hypothesis ($p \approx0.5$). 

\section{Conclusions}
We obtained relationships between the behavior of people from the words that they use by applying dimensional reduction to posts from the discussion site Reddit. We found the two most prominent dimensions that captured the greatest variation in word use across user dictionaries. We isolated nonlinear patterns in the reduced space by comparing the pronouns people use, the topics people discussed, and the entropy of their word use. By combining the set of differences among individuals into a few primary dimensions of variation we can identify how different behavioral traits are codependent. This strategy for integrating behavioral differences may have implications for the association of behavior with personality as well as other ways of characterizing individual differences. We found that users who discuss sports and games also used third-person male pronouns (`he') and first-person plural pronouns (`we') more often than other users in the population. Those users were distinct from groups of individuals who use second-person singular pronouns (`you') and those who discuss global politics. This is the beginning of a more complete characterization of the dimensions of differences in humans. 

We greatly appreciate the advice of Alfredo Morales and Irving Epstein.

\newpage

\section*{APPENDIX}

\subsection{Stop words}

``i", ``i'm", ``i've", ``i'll", ``i'd", ``you", ``your", ``you're", ``he", ``his", ``him", ``she", ``her", ``hers", ``we", ``us", ``our", ``ours", ``they", ``their", ``they're", ``them", ``my", ``me", ``it", ``its", ``it's", ``the", ``a", ``an", ``no", ``yes", ``not", ``good", ``bad", ``well", ``be", ``am", ``is", ``are", ``was", ``were", ``been", ``wasn't", ``aren't", ``weren't", ``being", ``isn't", ``have", ``has", ``had", ``hadn't", ``hasn't", ``haven't", ``having", ``do", ``does", ``did", ``doing", ``done", ``don't", ``doesn't", ``didn't", ``can", ``can't", ``couldn't", ``could", ``shouldn't", ``should", ``cannot", ``get", ``gets", ``got", ``gotten", ``getting", ``say", ``said", ``saying", ``will", ``would", ``wouldn't", ``go", ``goes", ``gone", ``went", ``going", ``come", ``coming", ``came", ``make", ``makes", ``made", ``know", ``knows", ``knowing", ``knew", ``known", ``take", ``takes", ``taking", ``took", ``taken", ``see", ``sees", ``saw", ``seeing", ``seen", ``look", ``looks", ``looked", ``looking", ``think", ``thinks", ``thought", ``thinking", ``use", ``uses", ``using", ``used", ``want", ``wants", ``wanted", ``wanting", ``give", ``gives", ``giving", ``gave", ``given", ``to", ``of", ``in", ``that", ``for", ``on", ``with", ``as", ``at", ``from", ``by", ``than", ``about", ``so", ``like", ``into", ``only", ``and", ``but", ``or", ``if", ``just", ``also", ``even", ``because", ``up", ``down", ``left", ``right", ``there", ``out", ``over", ``back", ``way", ``one", ``all", ``other", ``any", ``these", ``those", ``most", ``time", ``now", ``then", ``after", ``before", ``new", ``day", ``who", ``what", ``where", ``why", ``how", ``which", ``that", ``that's", ``this", ``too"

\subsection{Topics}
\begin{itemize}
\item ``Hockey'': `game', `team', `season', `hockey', `goal', `leafs', `adnan', `nhl', `tonight', `puck', `goalie', `win', `pp', `fans', `canucks', `hae', `net', `players', `ice', `player', `games', `jay', `kessel', `penalty', `teams', `goals', `fan', `playing', `gonna', `bruins', `period'
\item ``Video games'': `game', `damage', `play', `team', `player', `games', `level', `playing', `enemy', `items', `attack', `kill', `against', `weapons', `played', `armor', `skill', `weapon', `server', `win', `mana', `deck', `cards', `lane', `item', `hit', `card', `hp', `base', `gold', `power', `support', `map', `hero'
\item ``Global politics'': `people', `money', `against', `government', `world', `state', `police', `years', `country', `law', `power', `support', `system', `many', `war', `religion', `part', `american', `never', `america', `nothing', `evidence', `cops', `believe', `pay', `public', `rights', `states', `society', `political', `human', `children', `laws', `tax', `article', `kill', `religious'
\item ``actually'': `people', `more', `some', `when', `things', `very', `something', `much', `same', `actually', `point', `lot', `different', `need', `someone', `without', `thing', `anything', `may', `many', `work', `own', `might', `read', `reason', `kind', `while', `problem', `person', `etc', `find', `less', `still', `through', `understand', `enough', `case'
\end{itemize}

\end{document}